\documentclass[twocolumn,pra,showpacs]{revtex4}
\usepackage{amsmath}
\usepackage{amsfonts}
\begin{document}
\title{Quantization of four-dimensional Abelian  gravity}
\date{\today}
\author{Bogus{\l}aw Broda} \email[]{bobroda@uni.lodz.pl}
\author{Piotr Bronowski} \email[]{bronowski@gmail.com}
\author{Marcin Ostrowski}\email[]{ostrowsk@merlin.phys.uni.lodz.pl}
\author{Micha{\l} Szanecki}\email[]{michalszanecki@wp.pl}
\affiliation{Department of Theoretical Physics, University of
{\L}\'{o}dz Pomorska 149/153, 90-236 {\L}\'{o}dz, Poland}
\begin{abstract}
An abelian version of standard general relativity in the
Cartan--Palatini gauge-like formulation in four dimensions has
been introduced. Traditional canonical analysis utilizing
similarities to the akin Husain--Kucha{\v r} $SU(2)$ version of
gravity has been performed. The model has been next quantized in the
canonical path-integral Faddeev--Popov formalism yielding abelian
$BF$ theory.
\end{abstract}

\maketitle
\section*{Introduction}
\label{sec:intro}
Quantization of full four-dimensional gravity is a long-standing 
difficult problem of theoretical physics. There
are a lot of approaches to this problem, but two of them, loop
gravity and superstring theory seem to be the most promising. As is well-known the standard perturbative quantum-field-theory approach to
quantization of four-dimensional gravity fails because of
non-renormalizability. Quite another preliminary approach to
quantization program of gravity consists in investigation of
other, similar and simpler models. For example, gravity in lower
dimensions (in three \cite{witten1988dge}, or in two
\cite{grumiller2002dgt}) or other non-gravitational generally
covariant models like topological field theory (e.g.
\cite{witten1988tqf}). Proceeding in this spirit, we aim to
introduce and quantize a very naturally defined, generally
covariant, gravitation-like model in four dimensions. We will call
our model \textit{abelian gravity} because, as it follows from its
definition, it is a direct abelian analog of the standard general
relativity in the Cartan--Palatini gauge-like formulation. Our model
is $U(1)$ or abelian gravity in the same sense as the
Husain--Kucha\v{r} model is $SU(2)$ gravity \cite{husain1990gcn}.
We have also noticed that our model is akin to the ``toy theory'' introduced by
Oko{\l}\'ow \cite{okolow2006qdi}. It appears that the abelian
gravity has many interesting properties. It is, in a formal
hamiltonian sense, similar to the Husain--Kucha\v{r} $SU(2)$ gravity
and even, though in a more limited sense, to standard gravity. Because of
abelianity it is, in principle, simpler than the $SU(2)$ model but
the formal structure of constraints is more complex (the
constraints are not independent). In spite of the fact that our model
is abelian, it is non-linear and quite non-trivial. Actually, the
abelian gravity is a topological theory. It appears that the abelian
gravity can be explicitly quantized and thus solved. We shall
perform the standard Faddeev--Popov quantization procedure obtaining
the solution in a closed form.

In the first part of the paper (section \ref{sec:canonic}) we define our model and perform standard canonical analysis. Our derivation of the canonical
form is analogous to the one proposed by Husain and Kucha\v{r} in
the case of $SU(2)$ gravity \cite{husain1990gcn}. Qualitatively, the
corresponding expressions are simpler that in the $SU(2)$ case but
because of the non-invertibility of the set of one-forms $e^A_i$ the
situation is more complex and to maintain three-dimensional
symmetry we are forced to introduce a collection of
constraints which are not independent.

In the second part of our paper (section \ref{sec:quant}) we perform quantization of the abelian gravity in the formalism of path integrals. To this
end we use the most explicit and reliable canonical Faddeev--Popov
approach. In our case the whole procedure consists of several
steps. In the first step, we must throw out one of the dependent
constraints. In the second one, we perform a change of variables and
reexpress the path integral back in  an almost covariant form. Up to
that moment everything is gauge-independent. Next, we impose the
axial gauge-fixing condition for ordinary $U(1)$ gauge symmetry, and
``static'' (vanishing momenta) gauge condition for
diffeomorphic-like symmetry. The final step corresponds to the
original Faddeev--Popov trick consisting in exchange of gauges ---
trading off the canonical gauge condition for the covariant one.
Strictly speaking, in our case the trick is different because we do
not know the explicit covariant form of the gauge
(diffeomorphic-like) symmetry. Finally, we show that the abelian
gravity is equivalent to abelian $BF$ model.

\section{Canonical analysis}
\label{sec:canonic}
 The action of the four-dimensional abelian gravity is defined by
\begin{equation}
  S[A_\mu, e^A_\mu ]=\frac{1}{4}\int_{\mathcal{M}}{\mathrm{d}^4 x\,\epsilon^{\;\mu\nu\lambda\sigma}\epsilon_{AB}e^A_\mu e^B_\nu F_{\lambda\sigma}},
\label{eq:action}
\end{equation}
where the ``frame'' one-forms $e^A_\mu$ belong to the real
two-dimensional representation of $U(1),$ and
$F_{\mu\nu}=\partial_\mu A_\nu-\partial_\nu A_\mu$ is the $U(1)$
gauge curvature two-form. The Greek space-time indices run form $0$
to $3,$ i.e. $\mu,\nu,\lambda,\sigma=0,1,2,3,$ the capital Latin
internal indices run from $1$ to $2,$  i.e. $A,B=1,2.$ The alternating symbol
$\epsilon^{0123}=1,$ and $\epsilon_{12}=1.$

We observe that symbolically the action (\ref{eq:action}) is of the
form $\sim \int e\wedge e \wedge R,$ which is characteristic for
four-dimensional gravity in the Cartan--Palatini formalism, where
the gauge group $SO(1,3)$ has been replaced by $U(1).$ In
particular, the cosmological term $\sim \int e\wedge e\wedge e\wedge
e$ vanishes identically for $U(1).$

Denoting
\begin{equation}
 e^1_\mu\equiv B_\mu,\quad e^2_\mu\equiv C_\mu,
\end{equation}
we can rewrite the action (\ref{eq:action}) more conveniently as
\begin{eqnarray}
S[A_\mu,B_\mu,C_\mu]&=&\frac{1}{2}\int_{\mathcal{M}}{\mathrm{d}^4 x\, {\epsilon}^{\mu\nu\lambda\sigma}B_\mu C_\nu F_{\lambda\sigma}}=\nonumber\\
&=&\int_{{\mathcal M}={\mathbb R}\times\Sigma}{\mathrm{d}^4 x\,
{\epsilon}^{\mu\nu\lambda\sigma}B_\mu C_\nu\partial_\lambda
A_\sigma}.
\end{eqnarray}

Besides diffeomorphisms, there is a larger symmetry group than $SO(2)\cong U(1)$ present in the action. Namely, the full group $SL(2,{\mathbb R})$, acting locally according to formulas
\begin{subequations}
\begin{equation}
B'_\mu=aB_\mu+bC_\mu,
\label{eq:SL2Rb}
\end{equation}
\begin{equation}
C'_\mu=cB_\mu+dC_\mu,
\label{eq:SL2Rc}
\end{equation}
\end{subequations}
where
$\binom{\,a\,b\,}{\,c\,d\,} \in SL(2,{\mathbb R})$, leaves the action intact.
Since only a generator of the subgroup $SO(2)\cong U(1)$ appears among constraints, we can ignore the larger symmetry group in our further canonical quantization procedure.

We can now recast the action in the canonical form as
\begin{equation}
 \int_{\mathbb R}\mathrm{d}t\int_{\Sigma}\mathrm{d}^3 x\,\left(E^k \dot A_k +\frac{1}{2}\epsilon^{ijk}\epsilon_{AB}e^A_0e^B_i F_{jk}+A_0\partial_k E^k\right),
\label{eq:ActionSemiCov}
\end{equation}
where the momentum density
\begin{equation}
 E^k\equiv\frac{1}{2}\epsilon^{ijk}\epsilon_{AB}e^A_ie^B_j\equiv\epsilon^{ijk}B_iC_j
\end{equation}
or vectorially, simply $\vec E= \vec B\times \vec C,$ and surface
terms have been (temporarily)  discarded. Finally,
\begin{equation}
 S[A_i,E^i;N^i,N]=\int_{\mathbb R}\mathrm{d}t\int_{\Sigma}\mathrm{d}^3 x\,\left(E^i \dot A_i -N^i{\mathcal C}_i-N{\mathcal C}\right),
\label{eq:ActionWithConstraints}
\end{equation}
where $N,N^i$ are Lagrange multipliers and ${\mathcal C},\,{\mathcal
C}_i$ are constraints,
\begin{subequations}
\begin{equation}
 {\mathcal C}\equiv \partial_iE^i=0,
 \label{eq:Constraint1}
\end{equation}
\begin{equation}
{\mathcal C}_i\equiv E^jF_{ij}=0.
 \label{eq:Constraint2}
\end{equation}
\end{subequations}
Here, ${\mathcal C}$ is the ordinary abelian Gauss condition,
whereas ${\mathcal C}_i$ are (non-independent) spatial diffeomorphism
constraints. In contradistinction to general relativity there is no
hamiltonian constraint, and in contradistinction to the $SU(2)$
Husain--Kucha\v{r} model the constraints are not independent. We
could rewrite the diffeomorphism constraints (\ref{eq:Constraint2})
in a vector form
\begin{equation}
 {\mathcal C}_i\equiv\epsilon_{ijk}E^jF^k=0,\quad \vec {\mathcal C}\equiv\vec E\times \vec F=0,
\label{eq:DiffConstr}
\end{equation}
where the dual connection
$F^k\equiv\frac{1}{2}\epsilon^{ijk}F_{ij}.$ Obviously,
\begin{equation}
 E^i{\mathcal C}_i=0
\label{eq:constraintV1}
\end{equation}
identically, and therefore there are only two independent
constraints instead of three. In fact, three constraints would be
not admissible because we would get wrongly, negative number of
degrees of freedom, i.e. $3-(1+3)=-1.$ The twin condition
\begin{equation}
  F^i{\mathcal C}_i=0,
\label{eq:constraintV2}
\end{equation}
does not further limit the number of constraints because it reduces to the previous one upon the constraint condition itself. Namely, the ``vector'' constraint $\vec{\mathcal C}=0,$
 says that the vectors $\vec E$ and $\vec F$ should be parallel, $\vec E \parallel \vec F,$
 see Eq.~(\ref{eq:DiffConstr}), i.e.
\begin{equation}
  \vec E\propto \vec F,
\label{eq:paralelvec}
\end{equation}
and thus Eq.~(\ref{eq:constraintV2}) reduces to
Eq.~(\ref{eq:constraintV1}). In other words,
Eq.~(\ref{eq:paralelvec}) means that $\vec E=\alpha \vec F,$ and
consequently only one component of $\vec E$ is arbitrary, i.e.
$\alpha.$ Since, instead of six components (three for each of the
two vectors $\vec E$ and $\vec F$) we have only four, three for
$\vec F$ and one for $\alpha$ corresponding to the single degree of
freedom of $\vec E\quad (6-2=4).$ Therefore, we have effectively
exactly two independent constraints out of the three ${\mathcal
C}_i$ present in the action (\ref{eq:ActionWithConstraints}).

Now, we should check whether the canonical form of the action
(\ref{eq:ActionWithConstraints}) together with the constraints
(\ref{eq:Constraint1}) and (\ref{eq:Constraint2}) really corresponds
to the ``semi-covariant'' form (\ref{eq:ActionSemiCov}). First of
all, we observe that the first expected identification,
\begin{equation}
  N\equiv -A_0,
\end{equation}
directly follows from Eq.~(\ref{eq:ActionSemiCov}). Whereas the
``vector'' Lagrange multiplier is given by
\begin{equation}
  N^i=e^A_0e^i_A,
  \label{eq:LagrangeMultipl}
\end{equation}
where a new auxiliary field $e^i_A,$ the ``semi-inverse'' of
$e^A_i,$ has been introduced in Appendix. The use of $e^i_A$ yields
some redundancy, three non-independent constraints instead of two,
but the result assumes more symmetric form.

One can easily check that, as usually, the constraints
(\ref{eq:Constraint1}) and (\ref{eq:Constraint2}) are generators of
gauge transformation. Defining
\begin{equation}
  {\mathcal C}_\alpha\equiv \int \mathrm{d}^3x\, \alpha(x){\mathcal C}(x),\quad
{\mathcal C}_{\omega}\equiv\int\mathrm{d}^3x\,\omega^{i}(x){\mathcal
C}_i(x),
\end{equation}
we verify that
\begin{eqnarray}
 \delta_\alpha A_i&\equiv &\{{\mathcal C}_\alpha, A_i\}=\partial_i\alpha,
 \label{eq:Poisson1}\\
 \delta_\alpha E^i&\equiv &\{{\mathcal C}_\alpha, E^i\}=0
\label{eq:Poisson2}
\end{eqnarray}
and
\begin{eqnarray}
\delta_\omega A_i&\equiv &\{{\mathcal C}_\omega, A_i \}=\omega^jF_{ij},
\label{eq:Poisson3}
\\
\delta_\omega E^i&\equiv &\{{\mathcal
C}_\omega,E^i\}=\partial_j(\omega^iE^j-\omega^jE^i).
\label{eq:Poisson4}
\end{eqnarray}
Eqs.~(\ref{eq:Poisson1}) and (\ref{eq:Poisson2}) denote ordinary
abelian $U(1)$ gauge transformations, whereas (\ref{eq:Poisson3})
and (\ref{eq:Poisson4}) yield diffeomorphisms modulo $U(1)$ gauge
transformation and the Gauss condition, respectively. One could easily
redefine constraints to have diffeomorphisms in a ``pure'' form
(see \cite{husain1990gcn}) but it is unnecessary for our further purposes.

\section{Quantization}
\label{sec:quant}
We will use the most intuitive, easy and reliable
canonical Faddeev--Popov (path-integral) quantization procedure
\cite{faddeev1991gfi}. Since the original Faddeev--Popov procedure
is best suited to the case with independent constraints
\cite{slavnov1988qna}, one should discard, in our case, one of them,
${\mathcal C}_3,$ say. Intuitively, presence of a non-independent
constraint would correspond to a doubled Dirac delta and thus to a
trivial singularity in the path integral. Setting $N^3=0$, the path integral assumes the
following explicit form
\begin{eqnarray}
  Z&=&\int \exp\left[\frac{i}{\hbar} \int_{\mathbb R\times\Sigma}\mathrm{d}^4 x\,\,\,
\left(E^i\dot A_i+\epsilon_{ab}N^aE^bF^3 \right.\right.\nonumber \\
& &- \left.\left. \epsilon_{ab}N^aE^3F^b+
A_0\partial_iE^i\right)\right]\times\nonumber\\
& &\times G(A_i,E^i){\mathcal D}A_\mu{\mathcal D}E^i{\mathcal D}N^a,
\label{eq:PathInt1}
\end{eqnarray}
where $G(A_i,E^i)$ denotes, as yet unspecified, gauge part (gauge
conditions and Faddeev--Popov ghosts), and $a,b=1,2.$ Now, we
perform a minor change of variables,
\begin{equation}
  H_a\equiv \epsilon_{ab}N^bE^3,
  \label{eq:H}
\end{equation}
and consequently
\begin{equation}
 {\mathcal D} H_a=\epsilon_{ab}E^3{\mathcal D}N^b,\nonumber
\end{equation}
where we can treat $E^3$ as a constant. Then
\begin{eqnarray}
  Z&=&\int \exp\left[\frac{i}{\hbar} \int_{\mathbb R\times\Sigma}\mathrm{d}^4 x\,
\left(E^i\dot
A_i-\frac{H_aE^a}{E^3}F^3+H_aF^a \right.\right. \nonumber\\
& &\left.\left.+A_0\partial_iE^i\right)\right]
G(A_i,E^i){\mathcal D}A_\mu{\mathcal D}E^i\epsilon\frac{{\mathcal
D}H_a}{(E^3)^2}, \label{eq:PathInt2}
\end{eqnarray}
where $\epsilon=\pm 1$ is a (non-essential) regularization dependent
constant coming from the minus sign present in Eq.~(\ref{eq:H}).
Finally, we can reconstruct a (quasi) covariant form of $Z$
introducing the functional Dirac delta. Namely,
\begin{eqnarray}
   Z&=&\int \exp\left[\frac{i}{\hbar} \int_{\mathbb R\times\Sigma}\mathrm{d}^4 x\,
\left(E^iF_{0i}+H_iF^i\right)\right]\times\nonumber
\\
& &\times \delta(E^iH_i)G(A_i,E^i){\mathcal D}A_\mu{\mathcal
D}E^i\epsilon\frac{{\mathcal D}H_i}{E^3}, \label{eq:PathInt3}
\end{eqnarray}
where the functional integration with respect to an auxiliary (new)
variable $H_3$ yields the Eq.~(\ref{eq:PathInt2}) back. In
Eq.~(\ref{eq:PathInt3}) the primarily discarded surface term has
been recovered. Defining
\begin{equation}
  E^i\equiv\frac{1}{2}\epsilon^{0ijk}\sigma_{jk},\quad H_i\equiv\sigma_{0i}
\end{equation}
we obtain explicit ``covariantization'' of the path integral in
Eq.~(\ref{eq:PathInt3}),
\begin{equation}
  Z=\int\exp\left(\frac{i}{\hbar}\int_{\mathcal M}\sigma\wedge F\right)\delta(\sigma\wedge\sigma)\frac{\epsilon G}{\sigma_{12}}{\mathcal D} A{\mathcal D}\sigma.
\label{eq:PathInt4}
\end{equation}
Everything looks explicitly covariantly in Eq.~(\ref{eq:PathInt4})
except possibly the fraction in the measure. It is interesting to
note a similarity between Eq.~(\ref{eq:PathInt4}) and the ``toy
model'' presented in \cite{okolow2006qdi}.

To proceed further we should chose convenient gauge conditions. We
suggest the following canonical non-covariant gauge conditions

\begin{equation}
{\mathcal G}\equiv A_3=0,\quad {\mathcal G}^a\equiv E^a=0,
\label{eq:Gauge}
\end{equation}
which we could call ``axial-static'' ones. The gauge conditions
(\ref{eq:Gauge}) fulfill the necessary equalities
\cite{faddeev1991gfi}
\begin{equation}
 \{{\mathcal G},{\mathcal G}^a\}=\{{\mathcal G}^a,{\mathcal G}^b\}=0.
 \label{eq:Constr}
\end{equation}

One can easily calculate the corresponding Faddeev--Popov
determinant
\begin{eqnarray}
  \det M\equiv\left|\begin{array}{cc}
 \{{\mathcal C},{\mathcal G}\}&\{{\mathcal C},{\mathcal G}^b\}\\
\{{\mathcal C}_a,{\mathcal G}\}&\{{\mathcal C}_a,{\mathcal G}^b\}
\end{array}
 \right|=
\left|\begin{array}{cc}
 \partial_3&0\\
0&\delta^b_aE^3\partial_3
\end{array}
 \right|=\nonumber
\end{eqnarray}
\begin{eqnarray}
= |\det \partial_3|^3\prod_x [E^3(x)]^2.
\end{eqnarray}
Thus, the gauge-fixed form of the path integral (\ref{eq:PathInt4}) becomes
\begin{eqnarray}
  Z&=&\int\exp\left(\frac{i}{\hbar}\int\sigma\wedge F\right)\delta(\sigma\wedge\sigma)\epsilon\sigma_{12}
|\det \partial_3|^3\times\nonumber\\& &\times \delta(A_3)\delta(E^a){\mathcal D} A{\mathcal
D}\sigma. \label{eq:PathInt5}
\end{eqnarray}
Utilizing the gauge conditions for $E^a,$ we transform
Eq.~(\ref{eq:PathInt5}) to
\begin{eqnarray}
  Z=\int\exp\left[\frac{i}{\hbar}\int\mathrm{d}^4x\,(E^3F_{03}+H_iF^i)\right]\times
  \nonumber\\
  \times \delta(E^3H_3)\epsilon E^3|\det \partial_3|^3\delta(A_3){\mathcal
  D}A_\mu{\mathcal D}E^3{\mathcal D}H_i, \label{eq:PathInt6}
\end{eqnarray}
and finally, functionally integrating with respect to ${\mathcal D}H_3$ we obtain
\begin{eqnarray}
 Z &=& \int \exp\left[\frac{i}{\hbar}\int\mathrm{d}^4 x\, (E^3F_{03}+H_aF^a)\right]\times\nonumber\\
 & & \times \epsilon|\det \partial_3|^3\delta(A_3) {\mathcal D}A_\mu{\mathcal D}E^3{\mathcal D}H_a.
 \label{eq:PathInt7}
\end{eqnarray}
One could shorten the derivation of Eq.~(\ref{eq:PathInt7}) from
Eq.~(\ref{eq:PathInt1}) but we insist on keeping just this longer
derivation because of the excellent opportunity to present the
intermediate (quasi) covariant form of $Z$ given by
Eq.~(\ref{eq:PathInt4}).

Now we should perform the procedure analogous to the Faddeev--Popov
trick, consisting in replacing a non-covariant gauge by a covariant
one and yielding an explicitly covariant form of the path integral.
Actually, we do not invoke the original Faddeev--Popov trick because
we do not have at our disposal a covariant version of
(the ``reduced'' diffeomorphic) gauge transformations. This difficulty follows from the fact that the standard covariant form of diffeomorphic
transformations is implemented by four generators corresponding to
four constraints whereas we need only two of them. Instead,
we will show that canonically quantized four-dimensional
abelian $BF$ model assumes the form of Eq.~(\ref{eq:PathInt7}). Let
us recall the definition of the $BF$ theory \cite{blau1991tft}. The
action is defined by
\begin{eqnarray}
  S_{BF}[A_\mu,B_{\mu\nu}]\equiv\frac{1}{4}\int_{\mathcal M}\mathrm{d}^4x\,\epsilon^{\mu\nu\lambda\sigma}B_{\mu\nu}F_{\lambda\sigma},
\end{eqnarray}
where $F_{\mu\nu}=\partial_\mu A_\nu-\partial_\nu A_\mu$ as
earlier, and $B_{\mu\nu}$ is an abelian two-form.\\ Now,
\begin{eqnarray}
 S_{BF}&=&\frac{1}{2}\int \epsilon^{\mu\nu\lambda\sigma}B_{\mu\nu}\partial_\lambda A_\sigma=\nonumber\\
&=&\int_{\mathbb R}\mathrm{d}t\int_{\Sigma}\mathrm{d}^3x\left(E^i\dot A_i+ B_{0i}F^i+A_0\partial_iE^i\right),\quad
\end{eqnarray}
where, this time, the momentum density
\begin{equation}
 E^i\equiv\frac{1}{2}\epsilon^{ijk}B_{jk}.
\end{equation}
Then,
\begin{equation}
  S_{BF}[A_i,E^i;N_i,N]=\int_{\mathbb R\times \Sigma}\mathrm{d}^4x(E^i\dot A_i-N_i{\mathcal C}^i-N{\mathcal C}),
\end{equation}
where, in this model, the Lagrange multipliers are simply given by
\begin{equation}
 N\equiv -A_0,\quad N_i\equiv B_{i0},
\end{equation}
whereas the constraints
\begin{equation}
 {\mathcal C}\equiv\partial_i E^i=0,\quad {\mathcal C}^i\equiv F^i=0.
\end{equation}
In this model the constraints ${\mathcal C}^i$ are not independent
either because of the Bianchi identity
\begin{equation}
 \partial_i{\mathcal C}^i\equiv\partial_iF^i=0.
\end{equation}
Setting $N_3=0,$ as earlier, the path integral assumes the following
form
\begin{eqnarray}
 Z&=&\int \,\exp\left[\frac{i}{\hbar}\int_{\mathbb R\times\Sigma}\mathrm{d}^4x\,(E^i\dot A_i-N_aF^a+A_0\partial_iE^i)
 \right]\times\nonumber
\\
& &\times  G(A_i,E^i){\mathcal D}A_\mu{\mathcal D}E^i{\mathcal D}N_a,
\end{eqnarray}
where $G(A_i,E^i)$ denotes an appropriate gauge term to be specified.

To proceed further we chose the gauge conditions of the form
formally identical to (\ref{eq:Gauge}). Obviously,
Eqs.~(\ref{eq:Constr}) are also fulfilled. The Faddeev--Popov
determinant equals
\begin{equation}
 \det M\equiv\left|\begin{array}{cc}
 \partial_3&0\\
0&\epsilon_{ab}\partial_3
\end{array}
 \right|=\epsilon|\det\partial_3|^3,
\end{equation}
where $\epsilon$ is the regularization dependent sign introduced
earlier in our work. Now
\begin{eqnarray}
 Z=\int\,\exp\left[
\frac{i}{\hbar}\int\mathrm{d}^4x\, (E^iF_{0i}-N_aF^a)
\right]\times\nonumber
\\
\times\;\epsilon|\det \partial_3|^3\delta(A_3)\delta(E^a){\mathcal
D}A_\mu{\mathcal D}E^i{\mathcal D}N_a,
\end{eqnarray}
which is equal to Eq.~(\ref{eq:PathInt7}) if we set $H_a\equiv-N_a,$
and apply the second functional Dirac delta with respect to $E^a.$

\section{Conclusions}
\label{sec:concl}
In the previous section we have managed to show that the abelian gravity and abelian $BF$ theory,
upon an appropriate choice of non-covariant canonical gauge conditions in their canonical formulations, yield
the same path-integral formulas. Therefore, in spite of the fact that we do not know a covariant version
of the diffeomorphic-like gauge transformations, we can use the $BF$ model to covariantly quantize the abelian gravity. The procedure of quantization of $BF$ models in covariant gauges is well-known \cite{blau1991tft}.
It follows that the abelian gravity is a topological field theory, as it should, because
of the simple count of the number of degrees of freedom, i.e. $3-[1+(3-1)]=0.$
Although, from physical point of view, the abelian gravity does not resemble the standard gravity its
formal canonical structure does. Therefore, we believe that the model of abelian gravity could be of use
in theoretical physics applications.

Actually, the four-dimensional abelian gravity could be considered as a dual of four-dimensional abelian $BF$ model because it reproduces an equivalent system with other field content. Since in the abelian gravity the ``local base'' is spanned by the two vectors, $B_\mu$ and $C_\mu$, instead of four, one can interpret that model as a highly degenerated (topological) sector of the full (quantum) gravity. Therefore, one could speculate that what is known about the $BF$ system could be possibly used on gravitational side, but at the moment, we have not at our disposal any explicit example of this sort.

\begin{acknowledgments}
This work was supported in part by the Po\-lish Ministry
of Science and Higher Education Grant PBZ/MIN/008/P03/2003 and by the University of {\L}\'{o}d\'{z} grant.
\end{acknowledgments}

\section*{Appendix}
Let us define an auxiliary contravariant metric
\begin{equation}
g^{AB}\equiv\delta^{ij}e_i^Ae_j^B=\left(
\begin{array}{cc}
B^2& B\cdot C\\
B\cdot C & C^2
\end{array}
\right).
\end{equation}
Hence, the covariant metric $g_{AB},$ the (matrix) inverse of $g^{AB}$ is
\begin{equation}
g_{AB}=
\frac{1}{B^2C^2-(B\cdot C)^2}\left(
\begin{array}{cc}
C^2& -B\cdot C\\
-B\cdot C & B^2
\end{array}
\right).
\end{equation}
Now, we define the ``semi-inverse'' $e^i_A$ of $e^A_i,$ i.e.
\begin{equation}
e^i_A\equiv \delta^{ij}g_{AB}e^B_j.
\end{equation}
Actually,
\begin{equation}
e^A_i e^i_B=\delta^A_B.
\end{equation}
As an easy exercise, one can check that
\begin{equation}
-N^i{\mathcal C}_i=\frac{1}{2}\epsilon^{ijk}\epsilon_{AB}e^A_0e^B_iF_{jk},
\end{equation}
where $N^i$ and ${\mathcal C}_i$ are defined by Eq.~(\ref{eq:LagrangeMultipl}) and Eq.~(\ref{eq:Constraint2}), respectively.

\bibliography{literature}

\begin{thebibliography}{8}
\expandafter\ifx\csname natexlab\endcsname\relax\def\natexlab#1{#1}\fi
\expandafter\ifx\csname bibnamefont\endcsname\relax
  \def\bibnamefont#1{#1}\fi
\expandafter\ifx\csname bibfnamefont\endcsname\relax
  \def\bibfnamefont#1{#1}\fi
\expandafter\ifx\csname citenamefont\endcsname\relax
  \def\citenamefont#1{#1}\fi
\expandafter\ifx\csname url\endcsname\relax
  \def\url#1{\texttt{#1}}\fi
\expandafter\ifx\csname urlprefix\endcsname\relax\def\urlprefix{URL }\fi
\providecommand{\bibinfo}[2]{#2}
\providecommand{\eprint}[2][]{\url{#2}}

\bibitem[{\citenamefont{Witten}(1988{\natexlab{a}})}]{witten1988dge}
\bibinfo{author}{\bibfnamefont{E.}~\bibnamefont{Witten}},
  \bibinfo{journal}{Nuclear Physics B} \textbf{\bibinfo{volume}{311}},
  \bibinfo{pages}{46} (\bibinfo{year}{1988}{\natexlab{a}}).

\bibitem[{\citenamefont{Grumiller et~al.}(2002)\citenamefont{Grumiller, Kummer,
  and Vassilevich}}]{grumiller2002dgt}
\bibinfo{author}{\bibfnamefont{D.}~\bibnamefont{Grumiller}},
  \bibinfo{author}{\bibfnamefont{W.}~\bibnamefont{Kummer}}, \bibnamefont{and}
  \bibinfo{author}{\bibfnamefont{D.}~\bibnamefont{Vassilevich}},
  \bibinfo{journal}{Physics Reports} \textbf{\bibinfo{volume}{369}},
  \bibinfo{pages}{327} (\bibinfo{year}{2002}).

\bibitem[{\citenamefont{Witten}(1988{\natexlab{b}})}]{witten1988tqf}
\bibinfo{author}{\bibfnamefont{E.}~\bibnamefont{Witten}},
  \bibinfo{journal}{Communications in Mathematical Physics}
  \textbf{\bibinfo{volume}{117}}, \bibinfo{pages}{353}
  (\bibinfo{year}{1988}{\natexlab{b}}).

\bibitem[{\citenamefont{Husain and Kucha{\v r}}(1990)}]{husain1990gcn}
\bibinfo{author}{\bibfnamefont{V.}~\bibnamefont{Husain}} \bibnamefont{and}
  \bibinfo{author}{\bibfnamefont{K.}~\bibnamefont{Kucha{\v r}}},
  \bibinfo{journal}{Physical Review D} \textbf{\bibinfo{volume}{42}},
  \bibinfo{pages}{4070} (\bibinfo{year}{1990}).

\bibitem[{\citenamefont{Oko{\l}{\'o}w}(2006)}]{okolow2006qdi}
\bibinfo{author}{\bibfnamefont{A.}~\bibnamefont{Oko{\l}{\'o}w}},
  \bibinfo{journal}{Arxiv preprint gr-qc/0605138}  (\bibinfo{year}{2006}).

\bibitem[{\citenamefont{Faddeev and Slavnov}(1991)}]{faddeev1991gfi}
\bibinfo{author}{\bibfnamefont{L.}~\bibnamefont{Faddeev}} \bibnamefont{and}
  \bibinfo{author}{\bibfnamefont{A.}~\bibnamefont{Slavnov}},
  \emph{\bibinfo{title}{{Gauge fields, introduction to quantum theory}}}
  (\bibinfo{publisher}{Addison-Wesley}, \bibinfo{year}{1991}).

\bibitem[{\citenamefont{Slavnov and Frolov}(1988)}]{slavnov1988qna}
\bibinfo{author}{\bibfnamefont{A.}~\bibnamefont{Slavnov}} \bibnamefont{and}
  \bibinfo{author}{\bibfnamefont{S.}~\bibnamefont{Frolov}},
  \bibinfo{journal}{Theoretical and Mathematical Physics}
  \textbf{\bibinfo{volume}{75}}, \bibinfo{pages}{470} (\bibinfo{year}{1988}).

\bibitem[{\citenamefont{Blau and Thompson}(1991)}]{blau1991tft}
\bibinfo{author}{\bibfnamefont{M.}~\bibnamefont{Blau}} \bibnamefont{and}
  \bibinfo{author}{\bibfnamefont{G.}~\bibnamefont{Thompson}},
  \bibinfo{journal}{Ann. Phys} \textbf{\bibinfo{volume}{205}},
  \bibinfo{pages}{130} (\bibinfo{year}{1991}).

\end{thebibliography}
\end{document}